\documentclass[twocolumn,floatfix,eqsecnum,superscriptaddress,rmp]{revtex4}
\usepackage{graphicx}
\usepackage{amsmath}
\usepackage{bm}
\bibpunct{}{}{,}{s}{}{\textsuperscript{,}}
\newcommand{\be}{\begin{equation}}
\newcommand{\ee}{\end{equation}}

\begin{document}
\title{Thermo-electric properties of quantum point contacts}
\author{H. van Houten}
\affiliation{Philips Research Laboratories, 5600 JA Eindhoven, The Netherlands}
\author{L. W. Molenkamp}
\affiliation{Philips Research Laboratories, 5600 JA Eindhoven, The Netherlands}
\author{C. W. J. Beenakker}
\affiliation{Philips Research Laboratories, 5600 JA Eindhoven, The Netherlands}
\author{C. T. Foxon}
\affiliation{Philips Research Laboratories, Redhill, Surrey RH1 5HA, United Kingdom}
\begin{abstract}
The conductance, the thermal conductance, the thermopower and the
Peltier coefficient of a quantum point contact all exhibit quantum size effects. We
review and extend the theory of these effects. In addition, we review our
experimental work on the quantum oscillations in the thermopower, observed
using a current heating technique. New data are presented showing evidence for
quantum steps in the thermal conductance, and (less unequivocally) for quantum
oscillations in the Peltier coefficient. For these new experiments we have used a
quantum point contact as a miniature thermometer.\bigskip\\
{\tt Published in {\em Semiconductor Science \& Technology}
{\bf 7}, B215--B221 (1992).}
\end{abstract}
\maketitle

\tableofcontents

\section{\label{sec1} Introduction}

A quantum point contact is a short constriction of
variable width, comparable to the Fermi wavelength,
defined using a split-gate technique in a high-mobility
two-dimensional electron gas $(2\mathrm{D}\mathrm{E}\mathrm{G})$. Quantum point
contacts\cite{ref1,ref2} are best known for their quantized conductance at integer multiples of $2e^{2}/h$. For a general
review of quantum transport in semiconductor nanostructures see Ref.\ \cite{ref3}. The thermo-electric properties of
quantum point contacts have recently begun to be
explored as well.

The Landauer-B\"{u}ttiker formalism,\cite{ref4,ref5} which
treats electrical transport as a transmission problem
between reservoirs, has been generalized to thermal
transport and to thermo-electric cross-effects by Sivan
and Imry\cite{ref6} and by Butcher.\cite{ref7} Streda\cite{ref8} has
considered the specific problem of the thermopower $S$ of
a quantum point contact. He found that $S$ vanishes
whenever the conductance of the point contact is
quantized, and that it exhibits peaks between quantized
conductance plateaux. The magnitude of the peaks depends on the energy dependence of the transmission
probability $t(E)$ through the point contact. To the extent
that a quantum point contact behaves like an ideal
electron waveguide, $t(E)$ has a unit step-function energy
dependence. A somewhat more realistic model of a
quantum point contact --- introduced by B\"{u}ttiker\cite{ref9} --- is
to assume that the electrostatic potential has a saddle
shape. This particular model has also been used to
calculate the thermopower.\cite{ref10} The same theoretical
framework can be used to evaluate the thermal conductance $\kappa$ and the Peltier coefficient $\Pi$, which exhibit
quantum size effects similar to those in the conductance
and the thermopower, respectively. We review the theory
in section \ref{sec2}. For a discussion of thermo-electric effects in
different transport regimes, we refer to a recent article by
Ben-Jacob {\it et al}.\cite{ref11}

We have used a current heating technique\footnote{
Current heating has also been used by Gallagher {\it et al.}\cite{ref12} to study
fluctuations in the thermopower in the phase coherent diffusive
transport regime.
}
to observe the characteristic quantum size effects in the
thermo-electric properties of a quantum point contact.
Our previous work\cite{ref13,ref14} on the quantum oscillations in the
thermopower $S$ is reviewed in subsection \ref{sec3.1}.
Because of the sizable thermopower, a quantum point
contact can be used as a miniature `thermometer', to
probe the local temperature of the electron gas. We have
exploited this in our design of novel devices with multiple
quantum point contacts, with which we demonstrate
quantum steps in the thermal conductance $\kappa$ as well as
quantum oscillations in the Peltier coefficient of a
quantum point contact. The first results of these experiments are presented in subsections \ref{sec3.2} and \ref{sec3.3}. Concluding remarks are given in section \ref{sec4}.

\section{\label{sec2} Theoretical background}

\subsection{\label{sec2.1} Landauer-B\"{u}ttiker formalism of thermo-electricity}

The Landauer-B\"{u}ttiker formalism\cite{ref4,ref5} relates the
transport properties of a conductor to the transmission
probabilities between reservoirs that are in local equilibrium. Let us assume that only two such reservoirs are
present. In equilibrium, the reservoirs are at chemical
potential $E_{\mathrm{F}}$ and temperature $T$. In the regime of linear
response, the current $I$ and heat flow $Q$ are related to the
chemical potential difference $\Delta\mu$ and the temperature
difference $\Delta T$ by the constitutive equations\cite{ref15}
\be
\left(\begin{array}{c}
I\\
Q
\end{array}\right)=\left(\begin{array}{cc}
G & L\\
M & K
\end{array}\right)\left(\begin{array}{c}
\Delta\mu/e\\
\Delta T
\end{array}\right).   \label{eq1}
\ee
The thermo-electric coefficients $L$ and $M$ are related by
an Onsager relation, which in the absence of a magnetic
field is
\be
M=-LT. \label{eq2}
\ee
Equation (\ref{eq1}) is often re-expressed with the current $I$
rather than the electrochemical potential $\Delta\mu$ as an
independent variable,\cite{ref15}
\be
\left(\begin{array}{c}
\Delta\mu/e\\
Q
\end{array}\right)=\left(\begin{array}{cc}
R & S\\
\Pi & -\kappa
\end{array}\right)\left(\begin{array}{c}
I\\
\Delta T
\end{array}\right).\label{eq3}
\ee
The resistance $R$ is the reciprocal of the isothermal
conductance $G$. The thermopower $S$ is defined as
\be
S \equiv\left(\frac{\Delta\mu/e}{\Delta T}\right)_{I=0}=-L/G. \label{eq4}
\ee
The Peltier coefficient $\Pi$, defined as
\be
\Pi\equiv\left(\frac{Q}{I}\right)_{\Delta T=0}=M/G=ST,\label{eq5}
\ee
is proportional to the thermopower $S$ in view of the
Onsager relation (\ref{eq2}). Finally, the thermal conductance $\kappa$
is defined as
\be
\kappa\equiv-\left(\frac{Q}{\Delta T}\right)_{I=0}=-K\left(1+\frac{S^{2}GT}{K}\right). \label{eq6}
\ee

\begin{figure}
\centerline{\includegraphics[width=8cm]{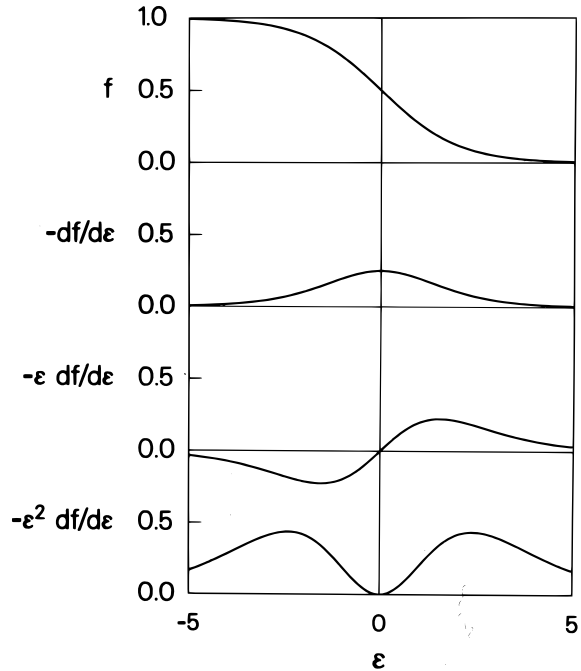}}
\caption{
From top to bottom: Fermi-Dirac distribution
function $f$, and $\epsilon^{m}\mathrm{d}f/\mathrm{d}\epsilon,$ for $m=0,1,2,$ as a function of
$\epsilon\equiv(E-E_{\mathrm{F}})/k_{\mathrm{B}}T$. These functions appear in expressions
(\ref{eq7},\ref{eq8},\ref{eq9}) for the thermo-electric coefficients.
\label{fig1}
}
\end{figure}

The thermo-electric coefficients are given in the Landauer-B\"{u}ttiker formalism by\cite{ref6,ref7}
\begin{eqnarray}
&&G=-\frac{2e^{2}}{h}\int_{0}^{\infty} dE\, \frac{\partial f}{\partial E}t(E),\label{eq7}\\
&&L=-\frac{2e^{2}}{h}\frac{k_{\mathrm{B}}}{e}\int_{0}^{\infty}dE\,\frac{\partial f}{\partial E}t(E)(E-E_{\mathrm{F}})/k_{\mathrm{B}}T, \label{eq8}\\
&&\frac{K}{T}=\frac{2e^{2}}{h}\left(\frac{k_{\mathrm{B}}}{e}\right)^{2}\int_{0}^{\infty}dE\,\frac{\partial f}{\partial E}t(E)[(E-E_{\mathrm{F}})/k_{\mathrm{B}}T]^{2}.\nonumber\\
&& \label{eq9}
\end{eqnarray}
These integrals are convolutions of $t(E)$, which characterizes the conductor, and a kernel of the form $\epsilon^{m}\mathrm{d}f/\mathrm{d}\epsilon$,
$m=0,1,2$, with $\epsilon\equiv(E-E_{\mathrm{F}})/k_{\mathrm{B}}T$, and $f$ the Fermi function
\be
f(\epsilon)=[\exp(\epsilon)+1]^{-1}. \label{eq10}
\ee
Plots of these kernels are given in figure \ref{fig1}.

Both $\mathrm{d}f/\mathrm{d}\epsilon$ and $\epsilon^{2}\mathrm{d}f/\mathrm{d}\epsilon$ are symmetric functions of $\epsilon$,
which is why the conductance, $G$, and the thermal
conductances $K$ and $\kappa$ are determined to first order by
$t(E_{\mathrm{F}})$. (The term within brackets in equation (\ref{eq6}) is usually
small.) In contrast, $\epsilon \mathrm{d}f/\mathrm{d}\epsilon$ is an anti-symmetric function
of $\epsilon$, so that the thermo-electric cross-coefficients $L,$ $S,$ $M$,
and $\Pi$ are determined mainly by the derivative $\mathrm{d}t(E)/\mathrm{d}E$
at $E=E_{\mathrm{F}}$. This is substantiated by a Sommerfeld expansion of the integrals (\ref{eq7},\ref{eq8},\ref{eq9}), valid for a smooth function
$t(E)$ to lowest order in $k_{\mathrm{B}}T/E_{\mathrm{F}}$,\cite{ref7}
\begin{eqnarray}
&&G\approx\frac{2e^{2}}{h}t(E_{\mathrm{F}}), \label{eq11}\\
&&L\approx\frac{2e^{2}}{h}L_{0}eT\left(\frac{\mathrm{d}t(E)}{\mathrm{d}E}\right)_{E=E_{\mathrm{F}}},\label{eq12}\\
&&K\displaystyle \approx-\frac{2e^{2}}{h}L_{0}Tt(E_{\mathrm{F}}), \label{eq13}
\end{eqnarray}
with $L_{0}\equiv(k_{\mathrm{B}}/e)^{2}\pi^{2}/3$ the Lorentz number. In this
approximation $K=-L_{0}TG$, so that for $S^{2}\ll L_{0}$ one
finds from (\ref{eq6}) the Wiedemann-Franz relation
\be
\kappa\approx L_{0}TG.
\label{eq14}
\ee
As discussed below, the thermo-electric coefficients of a quantum point contact may exhibit significant deviations from equations (\ref{eq11})--(\ref{eq14}). The inadequacy of the
Sommerfeld expansion is a consequence of the strong energy dependence of $t(E)$ near $E_{\mathrm{F}}$. In addition, $S^{2}\ll L_{0}$
does not hold for a quantum point contact close to pinch-off.

\subsection{\label{sec2.2} Quantum point contacts as ideal electron waveguides}

In this subsection we discuss the thermo-electric properties of a quantum point contact modelled as an ideal electron waveguide, matched perfectly to the reservoirs at entrance and exit. Such a waveguide has a transmission probability with step-function energy dependence
\be
t(E)= \sum_{n=1}^{\infty}\theta(E-E_{n}). \label{eq15}
\ee
The steps in $t(E)$ coincide with the threshold energies $E_{n}$
of the one-dimensional subbands or modes in the
quantum point contact.The integrals over the energy (\ref{eq7})
and (\ref{eq8}) determining the conductance and the thermopower
power can be evaluated analytically. By substitution of
(\ref{eq15}) into (\ref{eq7}), one finds for the conductance
\be
G=\frac{2e^{2}}{h}\sum_{n=1}^{\infty}f(\epsilon_{n})   \label{eq16}
\ee
with $\epsilon_{n}\equiv(E_{n}-E_{\mathrm{F}})/k_{\mathrm{B}}T$. This reduces to $G=(2e^{2}/h)N$ at
low temperatures ($N$ is the number of occupied subbands). Similarly, using the identity
\be
\int_{0}^{\infty}f\,dE=k_{\mathrm{B}}T\ln[1+\exp(E_{\mathrm{F}}/k_{\mathrm{B}}T)]   \label{eq17}
\ee
we find the exact result
\be
L=\frac{2e^{2}}{h}\frac{k_{\mathrm{B}}}{e}\sum_{n=1}^{\infty}[\ln(1+\mathrm{e}^{-\epsilon_{n}})+\epsilon_{n}(1+\mathrm{e}^{\epsilon_{n}})^{-1}].  \label{eq18}
\ee
The thermopower $S=-L/G$ and the Peltier coefficient
$\Pi=TS$ follow immediately from (\ref{eq16}) and (\ref{eq18}). At low
temperatures the thermopower vanishes, unless the
Fermi energy is within $k_{\mathrm{B}}T$ from a subband bottom. In
the limit $T=0$ one finds from (\ref{eq16}), (\ref{eq18}) that the maxima
are given by
\be
S=-\displaystyle \frac{k_{\mathrm{B}}}{e}\frac{\ln 2}{N-\frac{1}{2}},\;\; {\rm if}\;\; E_{\mathrm{F}}=E_{N};\;\;N>1. \label{eq19}
\ee
[Note that at $E_{\mathrm{F}}=E_{N}$ one also has $G=(2e^{2}/h)(N-\frac{1}{2})$.]
Equation (\ref{eq19}) was first obtained by Streda.\cite{ref8} For the
step-function model the width of the peaks in the thermopower as a function of $E_{\mathrm{F}}$ is of order $k_{\mathrm{B}}T$, at least in the
linear transport regime of small applied temperature
differences across the point contact $(\Delta T\ll T)$.

The thermopower of a quantum point contact with a
step-function $t(E)$ does not exhibit a peak near $E_{\mathrm{F}}=E_{1}$.
Instead, it follows from (\ref{eq16}) and (\ref{eq18}) that $-S$ increases
monotonically as $E_{\mathrm{F}}$ is reduced below $E_{1}$,
\be
S \approx-\frac{k_{\mathrm{B}}}{e}(1+\epsilon_{1}).   \label{eq20}
\ee
Note also that for $\epsilon_{1}\gg 1$, $S$ increases as $1/T$ as the
temperature is reduced. This result is probably not very
realistic. Indeed, for a saddle-shaped potential model of a 
quantum point contact we find instead in this regime
constant value which is proportional to $T$ (see subsection \ref{sec2.3}).

\begin{figure}
\centerline{\includegraphics[width=8cm]{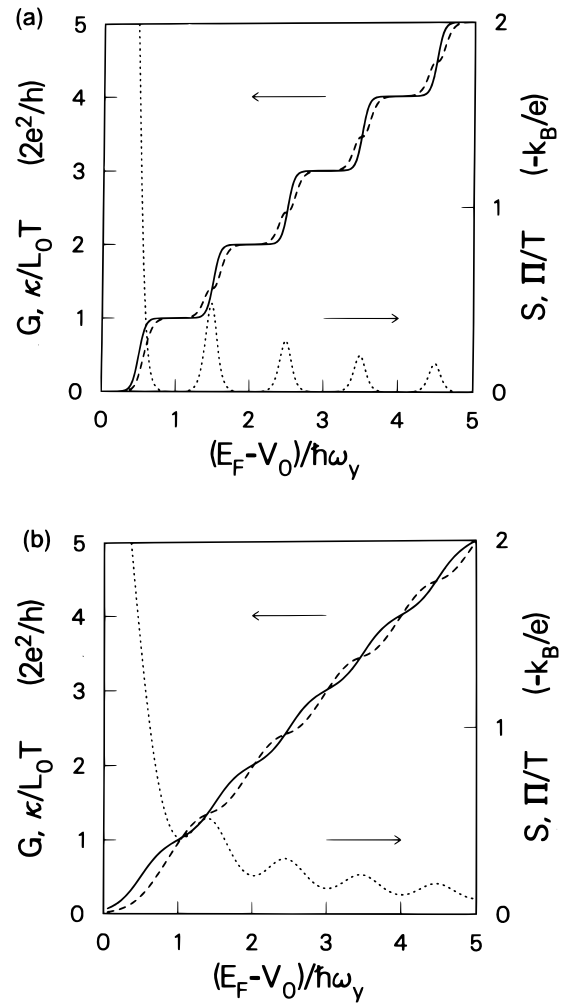}}
\caption{
Calculated conductance $G$ (full curve), thermal
conductance $\kappa/L_{0}T$ (broken curve), and the thermopower $S$ and 
Peltier coefficient $\Pi/T=S$ (same dotted curve) for a
quantum point contact with step function $t(E)$ as a function of Fermi energy at ({\em a}) 1~K and ({\em b}) 4~K.
The parameter used in the calculation is $\hbar\omega_{y}=2\,\mathrm{m}\mathrm{e}\mathrm{V}$.
\label{fig2}
}
\end{figure}

Plots of the thermo-electric coefficients as a function
of Fermi energy, calculated from (\ref{eq7}--\ref{eq9}) and (\ref{eq15}), are
given in figures \ref{fig2}(a) and \ref{fig2}(b), for $T=1\,\mathrm{K}$ and $T=4\,\mathrm{K}$
respectively. The values for $E_{n}$ are those for a parabolic
lateral confinement potential $V(y)=V_{0}+\frac{1}{2}m\omega^{2}_{y}y^{2}$ with 
$\hbar\omega_{y}=2.0\,\mathrm{m}\mathrm{e}\mathrm{V}$. We draw the following conclusions from 
these calculations.
\begin{enumerate}
\item
The temperature $T$ affects primarily the width of the
steps in $G$, and of the peaks in $S$, leaving the value of $G$ on
the plateaux, and the height of the peaks in $S$ essentially
unaffected.
\item
The thermal conductance $\kappa$ (divided by $L_{0}T$) exhibits
secondary plateaux near the steps in $G$, in violation of the
Wiedemann-Franz law. At 4~K the secondary plateaux
in $\kappa$ are even more pronounced than those in phase with
the plateaux in the conductance. These plateaux, which
apparently have not been noted previously, are due to the
bimodal shape of the kernel $\epsilon^{2}\mathrm{d}f/\mathrm{d}\epsilon$ (see figure \ref{fig1}).
\item
The coefficients $\kappa$ and $K$ differ from each other
whenever the thermopower $S$ does not vanish [cf.\ (\ref{eq6})]. We
have verified that this correction is usually negligible,
except in the vicinity of the first step in $G$.
\end{enumerate}

\subsection{\label{sec2.3} Saddle-shaped potential}

A more realistic model of a quantum point contact
should account for the rounding of the steps in $t(E)$. One
way to do this is to model the electrostatic potential
$V(x,$ $y)$ in the quantum point contact by a saddle-shaped
function\cite{ref9}
\be
V(x,y)=V_{0}-\frac{1}{2}m\omega_{x}^{2}x^{2}+\frac{1}{2}m\omega_{y}^{2}y^{2} \label{eq21}
\ee
where $V_{0}$ is the height of the saddle, $\omega_{x}$ characterizes the
curvature of the potential barrier in the constriction, and
$\omega_{y}$ the lateral confinement. The energies $E_{n}$ are given by
\be
E_{n}=V_{0}+(n-\frac{1}{2})\hbar\omega_{y}. \label{eq22}
\ee
The transmission probability is\cite{ref16}
\be
t(E)=\sum_{n=1}^{\infty}\left[1+\exp\left(\frac{-2\pi(E-E_{n})}{\hbar\omega_{x}}\right)\right]^{-1} \label{eq23}
\ee
Note that the step-function $t(E)$ is recovered in the limit
$\omega_{x}/\omega_{y}\rightarrow 0$.

\begin{figure}
\centerline{\includegraphics[width=8cm]{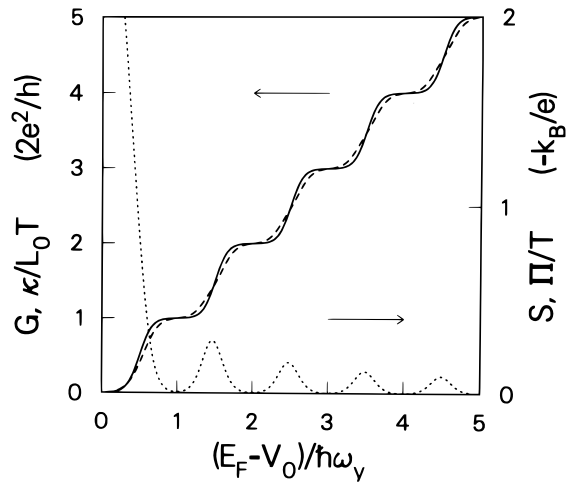}}
\caption{
Calculated conductance $G$ (full curve), thermal
conductance $\kappa/L_{0}T$ (broken curve), and the thermopower
$S$ and Peltier coefficient $\Pi/T=T$ (same dotted curve) for a
quantum point contact with a saddle shaped potential, as
a function of Fermi energy at 1 K. Parameters used in the calculation are $\hbar\omega_{\mathrm{y}}=2\,\mathrm{m}\mathrm{e}\mathrm{V}$, $\hbar\omega_{\mathrm{x}}=0.8\,\mathrm{m}\mathrm{e}\mathrm{V}$.
\label{fig3}
}
\end{figure}

To allow a comparison with the results in figure \ref{fig2} for
the step-function transmission probability, we have calculated the thermo-electric coefficients as a function of
Fermi energy from (\ref{eq7}--\ref{eq9}) and (\ref{eq23}), using the same value
of $2\,\mathrm{m}\mathrm{e}\mathrm{V}$ for the subband separation $\hbar\omega_{y}$, and taking
$\hbar\omega_{\mathrm{x}}\approx 0.8\,\mathrm{m}\mathrm{e}\mathrm{V}$ in order to reproduce the typically observed conductance step-widths at low temperatures. The
results at $T=4\,\mathrm{K}$ (not shown) were found to be identical
to those given in figure \ref{fig2}(b) for the step-function $t(E)$. At
$T=1\,\mathrm{K}$ there are some differences, however, as seen in
figure \ref{fig3}:
\begin{enumerate}
\item
The peak heights of the oscillations in the thermopower $S$ (or in the Peltier coefficient $\Pi$) are reduced by about a factor of two.
\item
The deviations from the Wiedemann-Franz law
$\kappa=L_{0}TG$ are much smaller. In particular, the secondary
plateau-like features (coinciding with the steps in $G$) are
absent.
\end{enumerate}

The behaviour of $S$ for $E_{\mathrm{F}}\ll E_{1}$ at low temperatures 
is qualitatively different from that discussed in subsection
\ref{sec2.2} for a step-function $t(E)$. Approximating
$t(E)\approx[1+\exp(2\pi(E_{1}-E)/\hbar\omega_{\mathrm{x}})]^{-1}$, and using the
Sommerfeld expansion results (\ref{eq11}) and (\ref{eq12}), we find that 
$S$ reaches an $E_{\mathrm{F}}$-independent value (not visible in figure \ref{fig3})
\be
S\approx-\frac{k_{\mathrm{B}}}{e}\frac{2\pi^{3}}{3}\frac{k_{\mathrm{B}}T}{\hbar\omega_{x}},\;\; E_{\mathrm{F}}\ll E_{1} \label{eq24}
\ee
which is proportional to $T$.

\section{\label{sec3} Experiments}

\subsection{\label{sec3.1} Thermopower}

\begin{figure}
\centerline{\includegraphics[width=8cm]{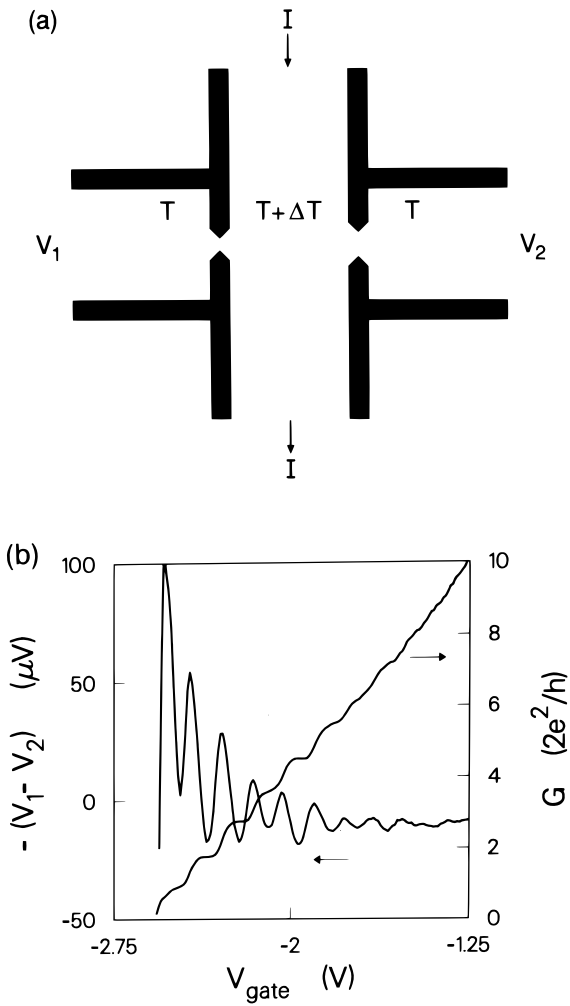}}
\caption{
(a) Schematic representation of the device
used to demonstrate quantum oscillations in the
thermopower of a quantum point contact by means of a
current heating technique. The channel has a width of 
4 $\mu \mathrm{m}$, and the two opposite quantum point contacts at its
boundaries are adjusted differently.
(b) Measured conductance and voltage $-(V_{1}-V_{2})$ as a function of the
gate voltage defining point contact 1, at a lattice temperature of 1.65 $\mathrm{K}$ and a current of 5 $\mu \mathrm{A}$. The gates
defining point contact 2 were kept at $-2.0$ V. 
\label{fig4}
}
\end{figure}

We have previously reported\cite{ref13,ref14} the observation of
quantum oscillations in the thermopower $S$ of a quantum 
point contact using a current heating technique. We 
review the main results here. The experimental arrangement is shown schematically in figure \ref{fig4}(a). By means of 
negatively biased split gates, a channel is defined in the
2DEG in a GaAs--AlGaAs heterostructure. A quantum
point contact is incorporated in each channel boundary. 
The point contacts 1 and 2 face each other, so that the 
voltage difference $V_{1}-V_{2}$ (measured using ohmic contacts attached to the $2\mathrm{D}\mathrm{E}\mathrm{G}$ regions behind the point
contacts) does not contain a contribution from the
voltage drop along the channel.

On passing a current $I$ through the channel, the
average kinetic energy of the electrons increases, because
of the dissipated power (equal to $(I/W_{\mathrm{c}\mathrm{h}})^{2}\rho$ per unit area,
for a channel of width $W_{\mathrm{c}\mathrm{h}}$ and resistivity $\rho$). We ignore
the net drift velocity acquired by the electron gas, and
assume that we can describe the non-equilibrium energy
distribution in the channel by a heated Fermi function at
temperature $T+\Delta T$. Since the point contacts are
operated as voltage probes, drawing no net current, the
temperature difference $\Delta T$ gives rise to a net
thermovoltage
\be
V_{1}-V_{2}=(S_{1}-S_{2})\Delta T. \label{eq25}
\ee
As dictated by the symmetry of the channel (see figure
\ref{fig4}(a)), this voltage difference vanishes unless the point
contacts are adjusted differently, so that they have
unequal thermopowers $S_{1}\neq S_{2}$.

A typical experimental result\cite{ref13} is shown in figure
\ref{fig4}(b). The gate voltage defining point contact 1 is scanned,
while that of point contact 2 is kept constant. In this way,
any change in the voltage difference $V_{1}-V_{2}$ is due to
variations in $S_{1}$. $(S_{2}$ is not entirely negligible, which is
why the trace for $-(V_{1}-V_{2})$ drops below zero in figure
\ref{fig4}(b).) Also shown is the conductance $G$ of point contact 1,
obtained from a separate measurement. For more
negative gate voltages, where the point contact resistance
exhibits quantized plateaux, we observe strong oscillations in $V_{1}-V_{2}$. The peaks occur at gate voltages where
$G$ changes stepwise because of a change in the number of
occupied $1\mathrm{D}$ subbands in point contact 1. These observations are a manifestation of the quantum oscillations in $S$ described in section \ref{sec2}.

A detailed comparison of the oscillations in figure \ref{fig4}(b)
with the ideal electron waveguide model (extended to 
the regime of finite thermovoltages and temperature 
differences) has been presented elsewhere.\cite{ref13} The decrease in amplitude of consecutive peaks is in agreement 
with equation (\ref{eq19}). We therefore only discuss the 
amplitude of the strong peak near $G=1.5(2e^{2}/h)$. The 
stepfunction transmission probability result (\ref{eq19}) predicts 
$S\sim-40\,\mu \mathrm{V}\mathrm{K}^{-1}$ for this peak, but a value 
$S\sim-20\,\mu \mathrm{V}\mathrm{K}^{-1}$ is probably more realistic (cf.\ figure \ref{fig3}). 
The measured value of about 50 $\mu \mathrm{V}$ for the amplitude of 
that peak thus indicates that the temperature of the
electron gas in the channel is $\Delta T\sim 2\,\mathrm{K}$ above the lattice
temperature $T=1.65$ K.

The increase in temperature $\Delta T$ is expected to be
related to the current in the channel by the heat balance
equation
\be
c_{\mathrm{v}}\Delta T=(I/W_{\mathrm{c}\mathrm{h}})^{2}\rho\tau_{\epsilon}   \label{eq26}
\ee
with $c_{\mathrm{v}}=(\pi^{2}/3)(k_{\mathrm{B}}T/E_{\mathrm{F}})n_{\mathrm{s}}k_{\mathrm{B}}$ the heat capacity per unit
area, $n_{\mathrm{s}}$ the electron density, and $\tau_{\epsilon}$ an energy relaxation
time associated with energy transfer from the electron gas
to the lattice. The symmetry of the geometry implies that
$V_{1}-V_{2}$ should be even in the current, and equation (\ref{eq26})
predicts more specifically that the thermovoltage difference $V_{1}-V_{2}\propto\Delta T$ should be proportional to $I^{2}$ --- at
least for small current densities. This is borne out by
experiment\cite{ref13,ref14} (not shown). Equation (\ref{eq26}) allows us
to determine the time $\tau_{\epsilon}$ from the experimental value
$\Delta T\sim 2$ K. Under the experimental conditions of figure
\ref{fig4}(b) we have $T=1.65\,\mathrm{K}$, $I=5\,\mu \mathrm{A}$, $W_{\mathrm{c}\mathrm{h}}=4\,\mu \mathrm{m}$, $\rho=20\,\Omega$.
We thus find $\tau_{\epsilon}\sim 10^{-10}\,\mathrm{s}$, which is not an unreasonable
value for the $2\mathrm{D}\mathrm{E}\mathrm{G}$ in GaAs--AlGaAs heterostructures at
helium temperatures.\cite{ref17}

The sudden decrease in $V_{1}-V_{2}$ beyond the last peak
(strong negative gate voltages) is not quite understood.
As discussed in section \ref{sec2}, the behaviour of $S$ in this
regime depends crucially on the details of the energy
dependence of $t(E)$.

\subsection{\label{sec3.2} Thermal conductance}

The sizable thermopower of a quantum point contact (up
to $-40\,\mu \mathrm{V}\mathrm{K}^{-1}$) suggests its possible use as a miniature
thermometer, suitable for local measurements of the
electron gas temperature. We have used this idea in an
experiment designed to demonstrate the quantum steps
in the thermal conductance of a second quantum point contact.

\begin{figure}
\centerline{\includegraphics[width=8cm]{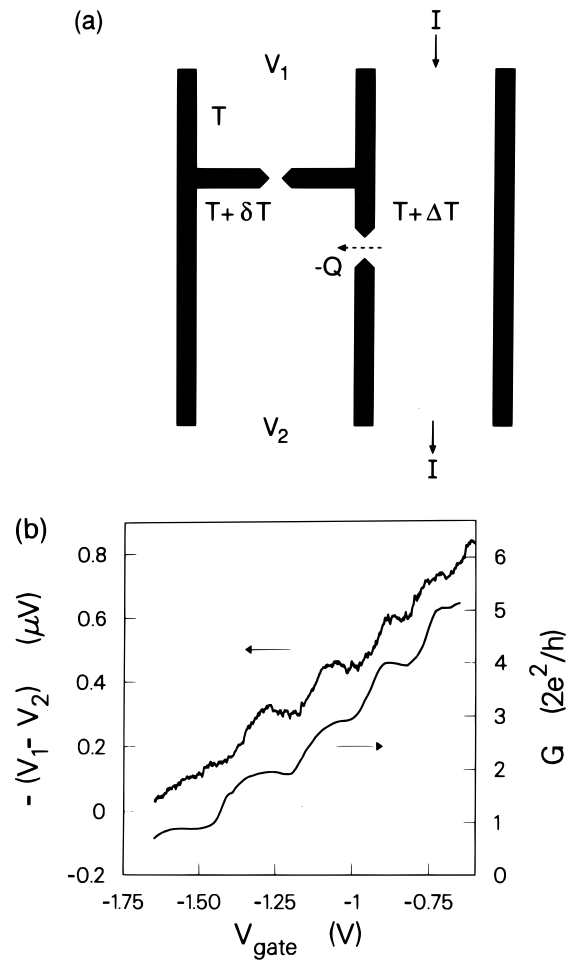}}
\caption{
({\it a}) Schematic representation of the device 
used to demonstrate quantum steps in the thermal
conductance of a quantum point contact, using another
point contact asa miniature thermometer.The main 
channel is $0.4\,\mu \mathrm{m}$ wide. ({\it b}) Measured conductance and 
{\scshape rms} of the second harmonic component of the contact 
voltage $V_{1}-V_{2}$ as a function of the gate voltage defining 
the point contact in the main channel boundary, at a lattice temperature of 1.4 $\mathrm{K}$ and an alternating current of
{\scshape rms} amplitude $0.6\,\mu \mathrm{A}$. The gates defining the other point
contact were kept at $-1.4\,\mathrm{V}$, so that its conductance is 
$G=1.5(2e^{2}/h)$.
\label{fig5}
}
\end{figure}

The geometry of the device is shown schematically in
figure \ref{fig5}(a).The main channel has a boundary containing
a quantum point contact. Using current heating, the
electron gas temperature in the channel is increased by
$\Delta T$, giving rise to a heat flow $Q$ through the point contact.
This causes a steady state temperature rise $\delta T$ of the
$2\mathrm{D}\mathrm{E}\mathrm{G}$ region behind the point contact (neglected in the
previous subsection), which we detect by a measurement
of the thermovoltage across a second point contact
situated in that region.

To increase the sensitivity of our experiment, we have
used a low-frequency $\mathrm{A}\mathrm{C}$ current to heat the electron gas
in the channel, and a lock-in detector tuned to the second
harmonic to measure the root-mean-square amplitude of
the thermovoltage $V_{1}-V_{2}$. The voltages on the gates
defining the second quantum point contact were adjusted
so that its conductance was $G=1.5(2e^{2}/h)$. Finally, we
applied a very weak magnetic field (15 $\mathrm{m}\mathrm{T}$) to avoid
detection of hot electrons on ballistic trajectories from
the first to the second point contact.

Figure \ref{fig5}(b) shows a plot of the measured thermovoltage as a function of the voltage on the gates defining the
point contact in the channel boundary, for a channel 
current of 0.6 $\mu \mathrm{A}$ ({\scshape rms} value). A sequence of plateaux is 
clearly visible, lining up with the quantized conductance
plateaux of the point contact. Since the measured ther-movoltage is directly proportional to $\delta T$, which in turn is 
proportional to the heat flow $Q$ through the point 
contact, this result is a demonstration of the expected 
quantum plateaux in the thermal conductance 
$\kappa\equiv-Q/\Delta T$ at zero net current [cf.\ (\ref{eq6})]. We have verified 
that the second-harmonic thermovoltage signal at fixed 
gate voltages is proportional to $I^{2}$, as expected. Let us
now see whether the magnitude of the effect can be 
accounted for as well.

To estimate the temperature increase $\delta T$ in the region
behind the point contact, we write the heat balance for
that region of area $A$ (valid if $\delta T\ll\Delta T$)
\be
\kappa\Delta T=c_{\mathrm{v}}A\delta T/\tau_{\epsilon}.   \label{eq27}
\ee
We assume that $A$ equals the square of the diffusion
length $(D\tau_{\epsilon})^{1/2}\sim 10\,\mu \mathrm{m}$, so that $\tau_{\epsilon}$ drops out of (\ref{eq27}).
On inserting the Wiedemann-Franz approximation
$\kappa\approx L_{0}TG$, with $G=N(2e^{2}/h)$, and using the expression
for the heat capacity per unit area given in the previous
subsection (with $n_{\mathrm{s}}=E_{\mathrm{F}}m/\pi h^{2}$), we find
\be
\frac{\delta T}{\Delta T}\approx N\frac{\hbar}{mD}.   \label{eq28}
\ee
In the experiment $D=1.4\,\mathrm{m}^{2}\mathrm{s}^{-1}$, so that at the $N=1$
plateau in the conductance, we have $\delta T/\Delta T\approx 1.2\times 10^{-3}$. The experimental curve in figure \ref{fig5}(b) was
obtained at a current density in the main channel of
$I/W_{\mathrm{c}\mathrm{h}}=1.2\,{\rm Am}^{-1}$, nearly equal to that used in the
thermopower experiment shown in figure \ref{fig4}(b). The analysis of the latter data indicated that $\Delta T\approx 2\,\mathrm{K}$ at this
current density. Consequently, $\delta T\approx 2\,\mathrm{m}\mathrm{K}$. The point
contact used as a thermometer (adjusted to
$G=1.5(2e^{2}/h))$ has $S\approx-20\,\mu \mathrm{V}\mathrm{K}^{-1}$ (see subsection \ref{sec2.3}),
so that we finally obtain $V_{1}-V_{2}\approx-0.05\,\mu \mathrm{V}$. The
measured value is larger (cf.\ the first plateau in figure \ref{fig5}(b)),
but only by a factor of two. All approximations considered, this is quite satisfactory.

\subsection{\label{sec3.3} Peltier effect}

\begin{figure}
\centerline{\includegraphics[width=8cm]{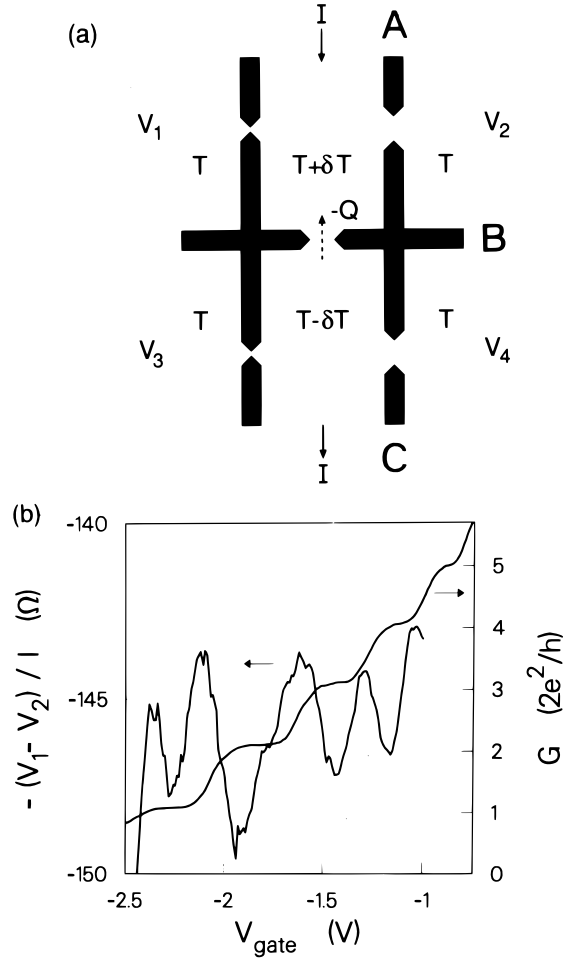}}
\caption{
(a) Schematic representation of the device
used to demonstrate quantum oscillations in the Peltier
coefficient of a quantum point contact. Arrows indicate
direction of positive flow. The main channel is $4\,\mu \mathrm{m}$ wide,
and the distance between the pairs of point contacts in its
boundaries is $20\,\mu \mathrm{m}$. (b) Measured conductance and
thermovoltage $-(V_{1}-V_{2})$ divided by the current $I$ as a
function of the voltage on gate B, defining the point
contact in the channel. The 1attice temperature is 1.6 $\mathrm{K}$
and the current is about 0.1 $\mu \mathrm{A}$ near $G=2e^{2}/h$. Gates
defining point contacts 1 and 3 were adiusted so that
their conductance was $G=1.5(2e^{2}/h)$. Gates A and $\mathrm{C}$
were unconnected.
\label{fig6}
}
\end{figure}

In this subsection we present preliminary results of an
experiment designed to observe the quantum oscillations
in the Peltier coefficient $\Pi$ of a quantum point contact.
The geometry of the experiment is shown schematically
in figure \ref{fig6}(a). A main channel, defined by split gates, is
separated in two parts by a barrier containing a point
contact. A positive current $I$ passed through this point
is accompanied by a negative Peltier heat flux
$Q=\Pi I$, giving rise to a (steady state) temperature rise $\delta T$
in the upper part of the channel, and to a temperature
drop $\delta T$ in the lower half. These temperature changes of
the electron gas can be detected by measuring the
thermovoltages across additional point contacts in the
channel boundaries --- least in principle.

One complication is that a total power $I^{2}/G$ is
dissipated due to the finite conductance $G$ of the
quantum point contact in the channel. This gives rise to a
temperature rise on both sides of the point contact. The
dissipated power is not equally distributed among the
$2\mathrm{D}\mathrm{E}\mathrm{G}$ regions on either side, and it is precisely this
imbalance which corresponds to the Peltier heat flow $\Pi I$.
We wish to detect only the temperature changes $\pm\delta T$
associated with the Peltier heat flow. This is accomplished by using an $\mathrm{A}\mathrm{C}$ current, and a lock-in detector tuned to the fundamental frequency to measure the
components linear in $I$ of the thermovoltages $(V_{1}-V_{2})$
and $(V_{3}-V_{4})$. The output voltage of the lock-in detector
is divided by the current, to obtain a signal linearly
proportional to the Peltier coefficient $\Pi$ of the point
contact in the channel. This signal, measured as a
function of the voltage on the gates defining that point
contact, should exhibit quantum oscillations, similar to
those seen in the thermopower $S$.

Unfortunately, our present sample design does not
allow us to do this without also affecting the thermopower of the point contacts used as thermometers. In
order to minimize this parasitic effect, we have scanned
only one of the gates (labelled B in figure \ref{fig6}(a)), and have
left the adjacent gates (A and C), which define the
reference point contacts, unconnected. The effect of gate
A on the remaining two thermometer point contacts is
negligible. A result obtained in this way (at $I\sim 0.1\,\mu \mathrm{A}$
and at $T=1.6\,\mathrm{K}$) is plotted in figure \ref{fig6}(b), together with a
trace of conductance versus gate voltage for the point
contact in the channel.

Oscillations in $-(V_{1}-V_{2})/I$ are clearly visible, of
amplitude up to $\approx 4\,\mathrm{V}$ A and with maxima aligned
with the steps between conductance plateaux. We interpret this signal as evidence for the oscillations in the
Peltier coefficient $\Pi$ (see below). However, the oscillations appear to be superimposed on a much larger
negative background signal. This signal (which we verified to be ohmic) is attributed to a series resistance
associated with the fact that gates A and C had to be
left unconnected, as mentioned above. The sum of the
contact resistance at the channel exit (estimated
at $(h/2e^{2})(\pi/2k_{\mathrm{F}}W_{\mathrm{c}\mathrm{h}})\approx 30\,\Omega)$ and the spreading resistance associated with current flowing to the wide $2\mathrm{D}\mathrm{E}\mathrm{G}$
regions of width $W_{\mathrm{w}\mathrm{i}\mathrm{d}\mathrm{e}}\approx 500\,\mu \mathrm{m}$ (estimated at
$\pi^{-1}\rho\ln(W_{\mathrm{w}\mathrm{i}\mathrm{d}\mathrm{e}}/W_{\mathrm{c}\mathrm{h}})\approx 30\,\Omega)$ is about $60\,\Omega$, which is of
about the correct magnitude to be able to account for the
background in figure \ref{fig6}(b). A new set of samples, designed
to avoid this background signal, are currently being
fabricated. ({\it Note} {\it added} {\it in} {\it proof}. Using these samples we
have indeed been able to observe the quantum oscillation
in $\Pi$ without such a background signal.\cite{ref19})

Let us now discuss the amplitude of the oscillations in
figure \ref{fig6}(b). To estimate $\delta T$, we use again the heat balance
equation, and find
\be
\delta T\approx\frac{\Pi  I}{c_{\mathrm{v}}D}.   \label{eq29}
\ee
Using the Onsager relation $\Pi=ST$, the estimated value
$S\approx-20\,\mu \mathrm{V}\mathrm{K}^{-1}$ for a quantum point contact adjusted
to $G=1.5(2e^{2}/h)$, and $T=1.65\,\mathrm{K}$, we deduce
$\delta T/I\approx 10^{4}\,\mathrm{K}\mathrm{A}^{-1}$. The resulting thermovoltage across
one of the thermometer point contacts (adjusted to
$G=1.5(2e^{2}/h)$ as well), normalized by $I$, is about
0.3 $\mathrm{VA}^{-1}$. This is ten times smaller than the experimentally observed amplitude of the corresponding oscillation
in figure \ref{fig6}(b). The origin of this discrepancy is not
understood.

\section{\label{sec4} Conclusions}

In conclusion, we have reviewed the theory of the
thermo-electric effects in a quantum point contact, and
our experiments on the quantum oscillations in the
thermopower. New data have been presented that --- for
the first time --- show evidence for the quantum steps in
the thermal conductance, and the quantum oscillations
in the Peltier coefficient. Our new experiments exploit
additional quantum point contacts as miniature thermometers. We have used this technique as well in an
experimental study of the effect of electron-electron
scattering on the ballistic mean free path.\cite{ref18} The
results for the thermo-electric transport coefficients presented here compare reasonably well with the theoretical
predictions. Further experiments as well as a more
reliable quantitative analysis would be desirable.

\acknowledgments

We acknowledge valuable contributions of M. J. P. Brugmans, R. Eppenga, M. A. A. Mabesoone, S. van Tuinen, and
Th.\ Gravier at various stages of the experiments, and
thank H. Buyk and C. E. Timmering for their technical
assistance. The support of M. F. H. Schuur\-mans is gratefully acknowledged. This research was partly funded
under the ESPRIT basic research action project 3133.

\end{document}